\documentclass[11pt]{article}
\usepackage[a4paper,margin=2.5cm]{geometry}
\usepackage{mathptmx}
\usepackage[expansion=false]{microtype}
\usepackage{amsmath,amssymb,amsthm}
\usepackage{graphicx}
\usepackage{booktabs}
\usepackage{multirow}
\usepackage[numbers,sort&compress]{natbib}
\usepackage[ruled,vlined,linesnumbered]{algorithm2e}
\usepackage{enumitem}
\usepackage{xcolor}
\usepackage[hidelinks]{hyperref}
\usepackage[capitalise,noabbrev]{cleveref}
\usepackage{setspace}
\newtheorem{theorem}{Theorem}

\theoremstyle{remark}

\newtheorem{finding}{Finding}

\newcommand{\G}[1]{\mathbb{G}_{#1}}
\newcommand{\Zr}{\mathbb{Z}_r^{*}}
\newcommand{\Ppub}{P_{\mathrm{pub}}}
\newcommand{\ID}{\mathit{ID}}
\newcommand{\sk}{\mathit{sk}}

\title{Authenticating IoT Data on Ethereum after Pectra and Fusaka:\\
A Cost Study of Signature Schemes, Batching and Anchoring\\
for Agricultural Supply Chain Finance}

\author{
Ming-Fong Sie\textsuperscript{a}, Jian-Hong Chen\textsuperscript{b}, Gwan-yen Lin\textsuperscript{c}, Yen-Shuo Chen\textsuperscript{c},
Shih-wei Liao\textsuperscript{b,*}
\\[1ex]
\small \textsuperscript{a}Master Program in Intelligent Computing and Big Data, Chung Yuan Christian University, Taoyuan, Taiwan\\
\small \textsuperscript{b}Department of Computer Science and Information Engineering, National Taiwan University, Taipei, Taiwan\\
\small \textsuperscript{c}Graduate Institute of Networking and Multimedia, National Taiwan University, Taipei, Taiwan\\
\small \textsuperscript{*}Corresponding author. E-mail: \href{mailto:liao@csie.ntu.edu.tw}{liao@csie.ntu.edu.tw}
}
\date{}

\begin{document}
\maketitle

\begin{abstract}
Blockchain-based supply chain finance can turn the sensor-recorded production history of smallholder farms into verifiable credit evidence, but only if every record is authenticated and anchored at a cost that scales to large device fleets. Ethereum's 2025 Pectra and Fusaka upgrades changed this cost landscape: they added precompiles for BLS12-381 and P-256, repriced calldata and MODEXP, and capped transaction gas at $2^{24}$. We organize the authentication and anchoring of IoT records into three trust levels, from posting a Merkle root through a trusted aggregator to verifying every device signature on-chain, and implement four replay-protected signature designs: ECDSA over secp256k1 and P-256, BLS aggregate signatures with on-chain hash-to-curve, and a pseudonymous identity-based signature with batch verification. Measured as signed transactions under the rules before and after both upgrades, the upgrades mainly changed which designs are feasible, not their cost. Batching and Merkle anchoring cut the cost of ECDSA authentication sixfold, to 8,963 gas per record, and secp256k1 ECDSA remains the cheapest authenticated design: P-256 costs 1.6 times as much, the identity-based design 2.7 times and BLS aggregation 3.9 times, dominated by on-chain hashing to the curve. One transaction carries at most 1,754 ECDSA-authenticated records but only 411 BLS-aggregated ones. On a Raspberry Pi Zero 2 W, pairing-based signing takes 5.7\,ms against 0.29\,ms for ECDSA. Pairing-based designs pay off only when the application needs their additional properties, and the identity-based design adds key escrow at the key generation centre. The benchmark harness is released for reproducibility.
\end{abstract}


\section{Introduction}
\label{sec:intro}

Smallholder farmers are the most upstream and the most financially fragile participants in agricultural supply chains. They rarely keep auditable records, hold little collateral and therefore struggle to obtain credit~\cite{tripoli2018emerging}. Blockchain-based supply chain systems make production and transaction data tamper-evident and shareable among farmers, distributors and lenders~\cite{kamilaris2019rise,wang2019understanding,caro2018blockchain,bhat2022agri}, and verifiable transparency is known to lower the cost of financing~\cite{chod2020financing}. If the production history of a farm can be verified, lenders can release staged payments when production milestones are reached.

The evidence comes from devices: in our deployment scenario, low-cost camera units photograph crops at fixed intervals, store the images off-chain and upload signed fingerprints. Every record must be authenticated, otherwise production histories can be fabricated, and the cost of doing so must scale to thousands of records per day. Where should authentication happen, with which signature scheme, and at what cost? Two recent Ethereum upgrades make these questions worth revisiting. Pectra (May 2025)~\cite{eip7600} added precompiles for the BLS12-381 pairing-friendly curve (EIP-2537~\cite{eip2537}) and increased the price of data-heavy calldata (EIP-7623~\cite{eip7623}). Fusaka (December 2025)~\cite{eip7607,ef2025fusaka} added a P-256 signature-verification precompile (EIP-7951~\cite{eip7951}), repriced MODEXP (EIP-7883~\cite{eip7883}), raised the default block gas limit to 60 million (EIP-7935~\cite{eip7935}) and capped the gas of every transaction at $2^{24}=16{,}777{,}216$ (EIP-7825~\cite{eip7825}). Pairing-based schemes that were impractical on-chain before are now possible, standard device signatures can be verified natively, and a single transaction can no longer absorb an arbitrarily large batch.

The literature on IoT blockchains rarely quantifies these trade-offs. Proposals typically adopt one scheme, often a pairing-based identity-based signature with batch verification, and evaluate its cryptographic operation counts off-chain, without comparing it to the simplest alternative of verifying standard ECDSA signatures with the precompiles every Ethereum client provides. The Multi-Layer Aggregate Verification (MLAV) design~\cite{wu2021mlav,sie2022secured}, in which two of the authors were involved, is an example: it used an identity-based batch-verification signature of the IBV family~\cite{zhang2008ibv} and reported only simulated, operation-count-based costs. Re-examining it, we found that the published signature is insecure, inheriting weaknesses already known for this family~\cite{lee2013toward}, which motivated us to measure, rather than assume, what each design costs.

\paragraph{Contributions.}
\begin{enumerate}[leftmargin=*]
\item \textbf{Design space and trust levels.} We organize the authentication and anchoring of IoT records into three trust levels (\cref{sec:space}): T0, a trusted aggregator posts a Merkle root; T1, it posts all record hashes; T2, a contract verifies every device signature. Within T2 we consider individual submission and batch submission with Merkle anchoring.
\item \textbf{Four implemented designs.} We implement ECDSA over secp256k1 (ecrecover) and P-256 (EIP-7951), BLS aggregate signatures with an RFC~9380-conformant on-chain hash-to-curve built from the EIP-2537 and MODEXP precompiles, and MLAV-IBS, a corrected pseudonymous identity-based signature with small-exponent and pseudonym-grouped batch verification (\cref{sec:designs}).
\item \textbf{Measurements under current rules.} We execute all designs as signed transactions on an EVM configured with the Osaka (Fusaka) rules, and under the Cancun and Prague rules that preceded them, and report gas per record, the largest batch per transaction under EIP-7825, block-level capacity, calldata, deployment and registration costs, and device- and aggregator-side timings (\cref{sec:method,sec:results}).
\item \textbf{Findings and guidelines.} We quantify when batching, anchoring and each signature scheme pay off, and derive guidelines for practitioners (\cref{sec:guidelines}). All code and data are released.
\end{enumerate}

\paragraph{Relation to prior work.} This article is a new study that builds on the MLAV architecture, introduced in a two-page conference paper~\cite{wu2021mlav} and described in full in a journal article~\cite{sie2022secured}, both co-authored by the first and last authors. It is not an extended version of either. The three-layer architecture (\cref{fig:platform}) and the supply chain finance workflow (\cref{fig:workflow}) are adapted from those works; both figures have been redrawn and revised. The signature analysed in \cref{sec:ibs} is the one specified in~\cite{sie2022secured}, and our analysis shows that it is insecure. The design-space framework, all four implementations, the MLAV-IBS scheme, the measurements and the analysis are new, and the results reported here supersede the simulated cost comparisons of~\cite{wu2021mlav,sie2022secured}.

\section{Background and related work}
\label{sec:related}

\paragraph{Blockchains for agri-food supply chains and finance.} Surveys document distributed ledgers for traceability, food safety and certification in agriculture~\cite{kamilaris2019rise,wang2019understanding}. Early systems combined RFID or sensor data with a ledger~\cite{tian2016agri,tian2017supply}; AgriBlockIoT demonstrated an implementation on Ethereum and Hyperledger~\cite{caro2018blockchain}; later work addresses interoperability between enterprise chains~\cite{bhat2022agri}. Chod et al.~\cite{chod2020financing} show that verifiable supply chain transparency lowers the cost of signalling operational quality to lenders. Comparable ledgers have been built for pharmaceutical supply chains~\cite{tseng2018governance}. These systems usually authenticate records by the transaction signature of the submitting account and do not analyze the cost of authenticating high-volume sensor data.

\paragraph{Blockchains for IoT.} Architectures by Dorri et al.~\cite{dorri2017blockchain} and Novo~\cite{novo2018blockchain} offload work from constrained devices to more capable nodes, as the aggregators in our setting do. Lee and Lee~\cite{lee2017blockchain} and Hu et al.~\cite{hu2019autonomous} used blockchains to distribute IoT firmware; the latter introduced pairing-based batch verification in this context. Rollups move execution off the base layer~\cite{thibault2022rollups}; Merkle anchoring is a simpler form of the same idea that commits to records without making them available.

\paragraph{Signature schemes and batch verification.} ECDSA over secp256k1 is Ethereum's native transaction signature and can be verified in a contract with the \texttt{ecrecover} precompile; P-256 is the curve supported by most secure elements, TPMs and WebAuthn authenticators. BLS signatures~\cite{boneh2001short} on pairing-friendly curves such as BLS12-381~\cite{bowe2017bls12} can be aggregated into a single signature~\cite{boneh2003aggregate,boneh2018compact}, at the price of hashing every message to the curve~\cite{rfc9380}. Identity-based signatures (IBS)~\cite{shamir1984identity,boneh2001ibe,cha2003ibs,hess2002ibs} remove certificates by deriving public keys from identities. Batch verification checks many signatures together~\cite{bellare1998batch,camenisch2007batch,ferrara2009practical}; without random weights it can be fooled~\cite{boyd2000attacking}. In vehicular networks, Zhang et al.'s IBV scheme~\cite{zhang2008ibv} started a large family of pseudonymous identity-based batch-verification schemes, and Lee and Lai~\cite{lee2013toward} pointed out that it is vulnerable to replay, lacks non-repudiation and accepts batches of invalid signatures. Hakuta et al.~\cite{hakuta2012batch} studied batches of only a few signatures.

\paragraph{Pairings on Ethereum.} Ethereum has offered BN254 pairing precompiles since 2017~\cite{eip197}; BN254 provides roughly 100 bits of security, whereas BLS12-381 targets about 128. EIP-2537 adds BLS12-381 addition, multi-scalar multiplication (MSM), pairing-check and map-to-curve precompiles~\cite{eip2537}. To our knowledge, no prior study has compared pairing-based and ECDSA-based authentication of IoT data on these precompiles under the Fusaka transaction cap.

\section{Application context}
\label{sec:context}

\begin{figure}[t]
\centering
\includegraphics[width=\linewidth]{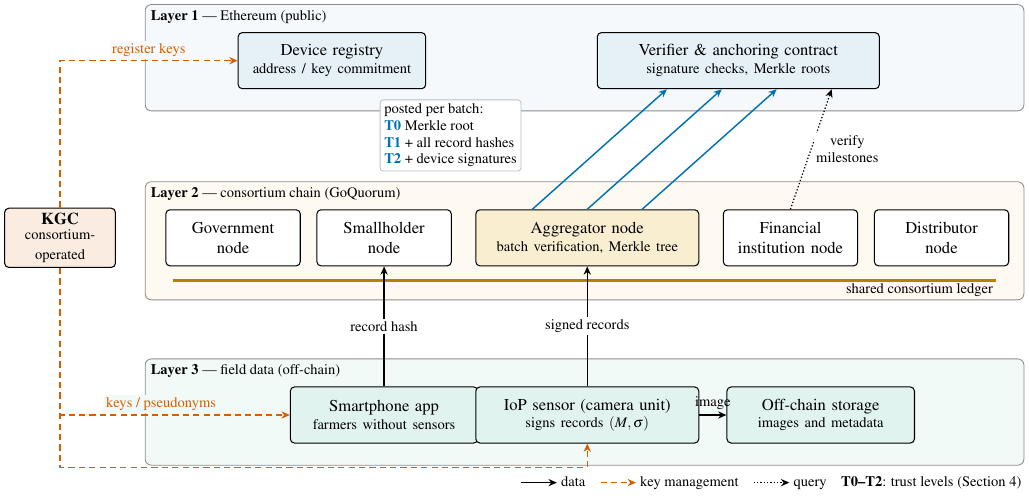}
\caption{MLAV platform (adapted from~\cite{wu2021mlav,sie2022secured}). Layer~3 collects field data off-chain, Layer~2 is a permissioned consortium chain whose aggregator verifies records in batches, and Layer~1 is Ethereum. What the aggregator posts per batch depends on the trust level T0--T2 of \cref{sec:space}; the KGC issues device keys and registers them on Layer~1.}
\label{fig:platform}
\end{figure}

\Cref{fig:platform} shows the MLAV platform, adapted from~\cite{wu2021mlav,sie2022secured}. A \emph{key generation centre} (KGC), operated by the consortium (for example the lender together with a public agency), registers farmers and their devices and can trace a device back to its owner in case of dispute. \emph{IoP sensors} (Raspberry Pi based camera units) photograph crops, store the images off-chain and sign a record $M=(CT,TS,Chk,Seed)$ of camera type, timestamp, image checksum and storage pointer (116 bytes in our encoding). Smallholders without sensors use a smartphone application. \emph{Aggregator nodes} of a permissioned consortium chain (GoQuorum~\cite{goquorum}) collect and verify records and anchor them on Ethereum. \emph{Financial institutions} grant loans and release staged payments, and \emph{distribution channels} sign purchase contracts and settle payments.

\begin{figure}[t]
\centering
\includegraphics[width=\linewidth]{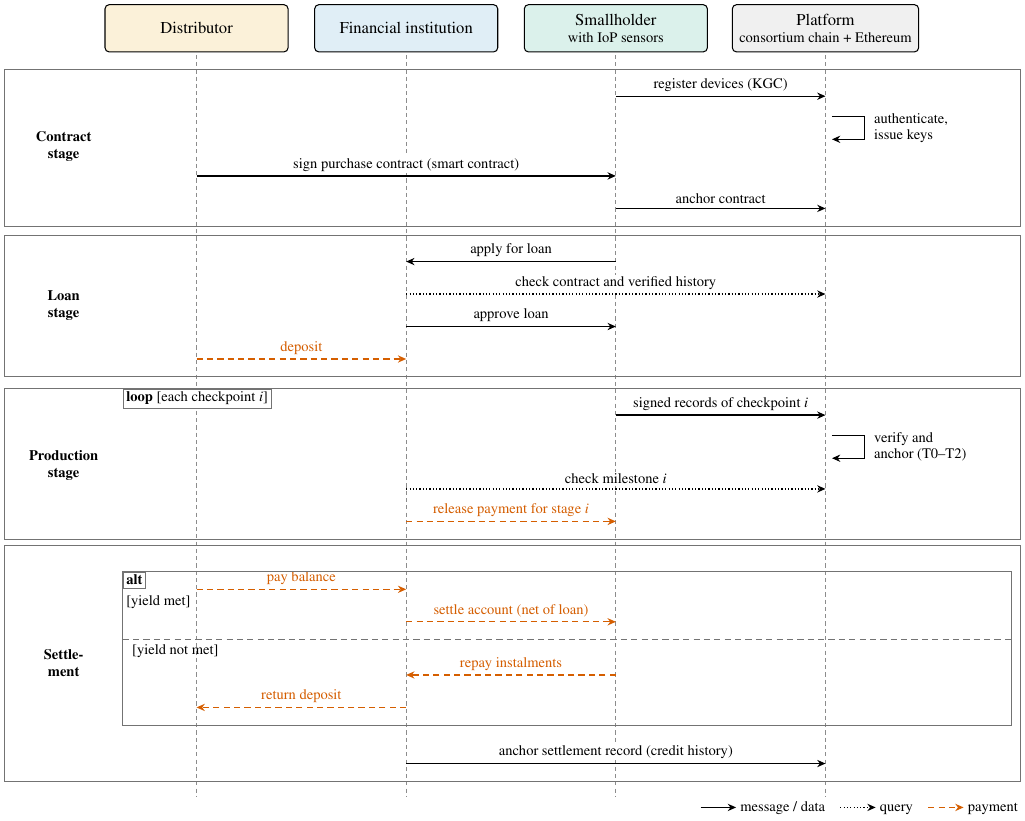}
\caption{Supply chain finance workflow of MLAV (adapted from~\cite{sie2022secured}), revised for this article. Signed production records are verified and anchored at each checkpoint before the staged payment is released; the settlement record becomes part of the farmer's credit history.}
\label{fig:workflow}
\end{figure}

\Cref{fig:workflow} shows the financing workflow. A distributor and a farmer sign a purchase contract; the farmer applies for a loan, and the lender approves it based on the contract and the farmer's verified production history, possibly with a deposit from the distributor. At each production checkpoint the farmer's devices upload signed records, which are verified and anchored, and the lender releases the payment for that stage. If the agreed yield is not met, the lender recovers the outstanding instalments and returns the deposit; otherwise the distributor pays the balance and the account is settled. Every step leaves a verifiable record that becomes part of the farmer's credit history.

\section{Design space}
\label{sec:space}

We distinguish three trust levels, according to what a verifier who trusts only Ethereum can check.

\begin{itemize}[leftmargin=*]
\item \textbf{T0 -- root only.} The aggregator verifies records off-chain, builds a Merkle tree~\cite{merkle1987digital} and posts the root. Anyone holding a record and its Merkle path can prove that the record was anchored at a given time, but the chain neither reveals which records exist nor attests that they were signed by registered devices. The aggregator is fully trusted for authenticity.
\item \textbf{T1 -- record hashes.} The aggregator posts all record hashes; the contract computes and stores the root. The set of anchored records becomes public and cannot later be extended or changed, which makes omissions and insertions detectable. Authenticity is still trusted to the aggregator.
\item \textbf{T2 -- on-chain authentication.} The contract verifies a device signature for every record before anchoring it, so that anchored records are authenticated even if the aggregator is malicious. Records may be submitted \emph{individually}, one transaction per record that stores the record hash, or in \emph{batches}, one transaction per batch that verifies all signatures and stores a Merkle root.
\end{itemize}

All T2 designs share a device registry maintained by the KGC: a contract storing, per device, either its address (secp256k1) or a 32-byte commitment to its public key, which the submitter provides in calldata. Without such a registry, any key could sign records; with it, identity-based signatures lose their main advantage of not requiring public-key distribution, a point we return to in \cref{sec:guidelines}.

\paragraph{Domain separation and replay protection.} In every T2 design the device signs the record together with a domain $d=(\mathit{chainid},\mathit{contract})$, so a signature is valid only for the chain and the contract it was made for. Signature verification alone does not stop a malicious aggregator from anchoring an old, validly signed record again, for instance an old photograph presented as evidence for a new payment milestone, and a batch is anchored as a single Merkle root, so the contract cannot recognise records it has already anchored. Each design therefore stores, per device, the timestamp of its last accepted record and accepts a record only if its timestamp is strictly larger. The slot is initialised at registration, so the same check also rejects unregistered devices, and each device can be registered only once, so the KGC cannot reset a device's timestamp or replace its key; a key change requires a new device entry. Within a batch, the first record of a device reads and updates the slot and later records of the same device find it warm, so the rule costs one storage update per device and batch (\cref{sec:results}). It requires the aggregator to submit each device's records in timestamp order: a record withheld until a later one of the same device has been anchored can no longer be anchored, a liveness limitation that a window of recently accepted timestamps would remove at a slightly higher cost.

\section{Authentication designs}
\label{sec:designs}

\subsection{ECDSA over secp256k1 (K1)}
The device signs $h=\mathrm{keccak256}(d\Vert M)$, where the domain $d$ prevents replay across chains and deployments. The contract rejects high-$s$ signatures (malleability), recovers the signer with \texttt{ecrecover} (3,000 gas), and checks and advances the signer's timestamp, whose slot doubles as the registry entry. A signature is 65 bytes.

\subsection{ECDSA over P-256 (R1)}
The device signs $h=\mathrm{SHA\text{-}256}(d\Vert M)$, as supported by secure elements. The contract checks the public key $(q_x,q_y)$ against the registry commitment and calls the EIP-7951 precompile (6,900 gas). In a batch, public keys are supplied once per device. A signature is 64 bytes.

\subsection{BLS aggregate signatures (BLS)}
Each device $j$ holds $\sk_j\in\Zr$ and $pk_j=\sk_jP_2\in\G2$, and signs $\sigma_i=\sk_j\,H(d\Vert c_j\Vert M_i)$, where $H$ is the RFC~9380 hash-to-curve suite \texttt{BLS12381G1\_XMD:SHA-256\_SSWU\_RO\_}~\cite{rfc9380} and $c_j$ the device's registry commitment. The aggregator sums all signatures of a batch into $\sigma=\sum_i\sigma_i$ (48 bytes compressed), so that calldata contains one signature per batch. The contract hashes every message to $\G1$, sums the hashes per device and checks
\begin{equation}
e(\sigma,-P_2)\cdot\prod_{j}e\Big(\sum_{i\in I_j}H(c_j\Vert M_i),\,pk_j\Big)=1,
\end{equation}
a pairing check with $m+1$ pairs for $m$ devices. Aggregation is secure for distinct messages~\cite{boneh2003aggregate}, which timestamps and checksums guarantee, and rogue-key attacks are prevented because keys are registered by the KGC. Our Solidity hash-to-curve implements \texttt{expand\_message\_xmd} with the SHA-256 precompile, reduces the two 64-byte outputs modulo $p$ with MODEXP and maps them with the EIP-2537 \texttt{MAP\_FP\_TO\_G1} precompile, which includes cofactor clearing; we verified that its output equals the reference implementation~\cite{noble}.

\subsection{Pseudonymous identity-based signature (IBS)}
\label{sec:ibs}

\paragraph{Why not the original MLAV signature.} The MLAV design~\cite{wu2021mlav}, specified in~\cite{sie2022secured}, used a signature of the IBV family~\cite{zhang2008ibv}: a device holds $PrK_1=c_1\,SID_1$ and $PrK_2=c_2\,H_1(SID_1\Vert SID_2)$, derived inside the device from the KGC's master keys $c_1,c_2$, and signs $\sigma=v\,PrK_1+w\,PrK_2$ with public scalars $v,w$ computed from the record; batches are verified by summing signatures. The signature is deterministic and linear in the two keys, so any two signatures of a device form a $2\times2$ linear system whose solution is $(PrK_1,PrK_2)$, enabling universal forgery for that device; plain-sum batching accepts pairs of invalid signatures whose errors cancel; and every device stores the master keys. Weaknesses of this kind have been reported for the IBV family~\cite{lee2013toward}; our released code reproduces both attacks against the scheme as specified in~\cite{sie2022secured}. The scheme of~\cite{sie2022secured} therefore does not provide unforgeability and should not be used; we replace it.

\paragraph{MLAV-IBS.} We use a pseudonymous variant of the Cha--Cheon IBS~\cite{cha2003ibs} over the Type-3 pairing $e:\G1\times\G2\to\G{T}$ of BLS12-381 (\cref{alg:ibs}). The KGC holds $s$ and publishes $\Ppub{}_{,1}=sP_1$ and $\Ppub{}_{,2}=sP_2$. A pseudonym $\ID=SID_1\Vert SID_2$ with $SID_1=tP_1$ and $SID_2=RID\oplus H_0(s\,SID_1)$ hides the real identity $RID$ from everyone but the KGC; the device stores only $\sk_{\ID}=s\,H_1(\ID)$. A signature is $(U,V)=(\rho Q_{\ID},(\rho+h)\sk_{\ID})$ with fresh $\rho$ and $h=H_2(\ID,M,U)$, where $M$ includes the domain $d$ as in all designs, and is valid if $e(V,P_2)\,e(U+hQ_{\ID},-\Ppub{}_{,2})=1$. Unforgeability follows from the Cha--Cheon proof~\cite{cha2003ibs} adapted to Type-3 pairings, under the assumption that computing $abP_1$ from $(P_1,aP_1,bP_1,P_2,aP_2)$ is hard; publishing $\Ppub{}_{,1}$ lets the reduction simulate extraction queries (\cref{app:proof}).

\begin{algorithm}[t]
\caption{MLAV-IBS}
\label{alg:ibs}
\SetKwProg{Fn}{}{:}{}
\Fn{\textsf{Setup} \textnormal{(KGC)}}{ $s\xleftarrow{\$}\Zr$;\quad \Return $\Ppub{}_{,1}=sP_1$, $\Ppub{}_{,2}=sP_2$\; }
\Fn{\textsf{Extract}$(RID)$ \textnormal{(KGC)}}{
  $t\xleftarrow{\$}\Zr$;\ $SID_1\gets tP_1$;\ $SID_2\gets RID\oplus H_0(s\,SID_1)$;\ $\ID\gets SID_1\Vert SID_2$\;
  $Q_{\ID}\gets H_1(\ID)$;\quad \Return $\sk_{\ID}=s\,Q_{\ID}$ \tcp*{only $\sk_{\ID}$ is stored on the device}
}
\Fn{\textsf{Sign}$(\sk_{\ID},M)$}{ $\rho\xleftarrow{\$}\Zr$;\ $U\gets\rho Q_{\ID}$;\ $h\gets H_2(\ID,M,U)$;\ \Return $(U,\,V=(\rho+h)\sk_{\ID})$\; }
\Fn{\textsf{Verify}$(\ID,M,U,V)$}{ \Return $e(V,P_2)\cdot e(U+H_2(\ID,M,U)\,Q_{\ID},\,-\Ppub{}_{,2})\overset{?}{=}1$\; }
\Fn{\textsf{Trace}$(\ID)$ \textnormal{(KGC)}}{ \Return $RID=SID_2\oplus H_0(s\,SID_1)$\; }
\end{algorithm}

\paragraph{Batch verification.} With random $\ell$-bit weights $\delta_i$, the small-exponent test~\cite{bellare1998batch} (B2) checks
\begin{equation}
e\Big(\sum_i\delta_iV_i,\,P_2\Big)\cdot e\Big(\sum_i\delta_iU_i+\sum_i\delta_ih_i\,Q_{\ID_i},\,-\Ppub{}_{,2}\Big)=1 .
\label{eq:b2}
\end{equation}
Because $Q_{\ID}$ depends only on the pseudonym, the identity terms can be collected per pseudonym, $\sum_i\delta_ih_iQ_{\ID_i}=\sum_{j}\big(\sum_{i\in I_j}\delta_ih_i\big)Q_j$, which we call \emph{pseudonym-grouped} batching (B3). It shrinks the second MSM from $2n$ to $n+m$ points and the number of evaluations of $H_1$ from $n$ to $m$. B3 is an algebraic regrouping of the small-exponent test, not a new verification technique, and its saving rests on many records sharing a pseudonym. Records signed under one pseudonym are linkable to each other, though not to the device's real identity. A device that changes its pseudonym for every record to make its records unlinkable gains nothing from grouping ($m=n$) and needs a new on-chain registration per pseudonym (\cref{sec:results}). Grouping and unlinkability between records are therefore alternatives. In the supply chain finance setting, the records of a farm are meant to be linked to its loan, so pseudonyms are long-lived and grouping applies; where records must be unlinkable, B2 is the relevant baseline.

\begin{theorem}[Batch soundness]
\label{thm:batch}
Let all $U_i,V_i\in\G1$ and let the $\delta_i$ be independent and uniform over a set $\Delta$ of integers that are distinct modulo $r$. If some signature in the batch is invalid, B2 and B3 accept with probability at most $1/|\Delta|$.
\end{theorem}
\begin{proof}
Let $\gamma_i=e(V_i,P_2)\,e(U_i+h_iQ_{\ID_i},-\Ppub{}_{,2})$, so that signature $i$ is valid iff $\gamma_i=1$. By bilinearity both checks compute $\prod_i\gamma_i^{\delta_i}$. Writing $\gamma_i=g^{\alpha_i}$ for a generator $g$ of $\G{T}$, acceptance means $\sum_i\alpha_i\delta_i\equiv0\pmod r$. If $\alpha_j\neq0$, then for any fixed $\{\delta_i\}_{i\neq j}$ exactly one residue of $\delta_j$ satisfies the congruence, which $\delta_j$ hits with probability at most $1/|\Delta|$.
\end{proof}

Off-chain, we use 64-bit odd weights (error $2^{-63}$). A contract has no private randomness, so we derive 128-bit weights as $\delta_i=\lfloor\mathrm{keccak256}(\mathrm{keccak256}(\mathit{calldata}),i)/2^{128}\rfloor$ in the Fiat--Shamir style~\cite{fiat1986prove}; an adversary who tries $q$ batches succeeds with probability at most $q/2^{128}$. Subgroup membership, required by \cref{thm:batch}, is checked by our deserializer and by the EIP-2537 MSM and pairing precompiles. On-chain, the KGC registers $c_{\ID}=\mathrm{keccak256}(Q_{\ID})$, submitters provide $Q_{\ID}$, and $H_2$ is keccak256 modulo $r$ over the domain $d$, $c_{\ID}$, $M$ and $U$.

\section{Methodology}
\label{sec:method}

\paragraph{On-chain.} Contracts are written in Solidity 0.8.28 (optimizer, 200 runs, via-IR) and executed as signed EIP-1559 transactions on the EthereumJS VM 10.1.3~\cite{ethereumjs} configured with the Osaka hardfork, which enforces EIP-2537, EIP-7623, EIP-7825, EIP-7883 and EIP-7951. We report the total gas of each transaction, including the 21,000 intrinsic gas and calldata, in a block with a non-zero block number so that every anchoring write is a fresh zero-to-non-zero storage write. Signatures and test data are produced with independent JavaScript implementations~\cite{noble}. Individual submission is measured as the mean of five transactions. Batches contain $n$ records from $m=\min(n,10)$ devices, round-robin. The largest batch under the EIP-7825 cap is found by bisection with actual transactions. Gas figures are deterministic.

\paragraph{Off-chain.} Device signing and aggregator verification are measured in Python 3.12 on a single virtual core of an Intel Xeon processor at 2.10\,GHz (Linux, glibc 2.39), the reference machine: ECDSA with \texttt{libsecp256k1}~\cite{libsecp256k1} (via \texttt{coincurve} 21.0) and OpenSSL 3.5 (via \texttt{cryptography} 46.0), and BLS and MLAV-IBS with the Rust \texttt{arkworks} library~\cite{arkworks} (\texttt{py\_arkworks\_bls12381} 0.5.0). We report medians of 7 repetitions (140 for signing). To check that the results do not depend on the processor, we repeated all off-chain measurements on GitHub-hosted runners with an ARM64 (Arm Neoverse-N2) and an x86-64 (AMD EPYC 9V74) processor (\cref{tab:platforms}): signing times stay within 18\% and per-record verification times within 31\% of the reference machine, and the ranking of the designs is unchanged. These are server-class cores, so we also measured two Raspberry Pi boards, recording each board's clock and throttling state before and after every run: every timing in \cref{tab:platforms} is 4.4 to 7.3 times slower on a Raspberry Pi 4 than on the reference machine and 11 to 16 times slower on a Raspberry Pi Zero 2 W, the class of board used in the IoP sensors, and the ranking of the designs is again unchanged. To measure aggregator throughput on several cores, worker processes verify independent, pre-signed batches of 1,000 records concurrently, and throughput is the total number of records divided by the slowest worker's time. On the Raspberry Pi boards every measurement was repeated 10 times, with a pause between runs, and we report medians across runs together with the largest coefficient of variation.

\paragraph{Correctness checks.} For every design, valid batches are accepted and batches with a tampered record are rejected; the on-chain hash-to-curve matches RFC~9380; and the MLAV-IBS verifiers reject the cancellation batch that the plain-sum verifier accepts, both off-chain and on-chain.

\section{Results}
\label{sec:results}

\subsection{Anchoring without authentication (T0, T1)}

Posting only a Merkle root (T0) costs 45,507 gas per transaction regardless of $n$, i.e.\ 91 gas per record for $n=500$. Posting all record hashes and computing the root on-chain (T1) costs 659,610 gas for 500 records (1,319 gas per record), and 1,298,080 gas for 1,000 records. Storing each record hash in its own storage slot instead costs 11,390,122 gas for 500 records (22,780 per record), 17 times more than T1, and 1,000 per-slot writes exceed the transaction cap. Merkle anchoring is therefore the single largest saving available, and it is independent of the signature scheme.

\subsection{On-chain authentication (T2)}

\begin{table}[t]
\centering
\caption{Gas per record under Osaka rules ($m=\min(n,10)$ devices). ``Individual'' is one transaction per record that verifies its signature and stores the record hash; batches verify all signatures and store one Merkle root. The last column is the largest batch that fits under the EIP-7825 cap of $2^{24}$ gas.}
\label{tab:gas}
\small\setlength{\tabcolsep}{4pt}
\begin{tabular}{@{}lrrrrrr@{}}
\toprule
Design & Individual & $n=10$ & $n=100$ & $n=200$ & $n=500$ & Max.\ $n$ \\
\midrule
T0 root only (no authentication)      & --        & 4{,}548   & 455      & --       & 91       & -- \\
T1 record hashes (no authentication)  & --        & 5{,}762   & 1{,}644  & --       & 1{,}319  & -- \\
\midrule
K1 ECDSA secp256k1                    & \textbf{55{,}475} & 17{,}896 & 9{,}492 & 9{,}078 & \textbf{8{,}963} & \textbf{1{,}754} \\
R1 ECDSA P-256                        & 63{,}788  & 27{,}066 & 15{,}129 & 14{,}580 & 14{,}542 & 1{,}105 \\
BLS aggregate                         & 187{,}118 & 82{,}455 & 34{,}927 & 35{,}230 & --      & 411 \\
IBS small-exponent (B2)               & 176{,}943 & 58{,}480 & 31{,}242 & 30{,}296 & 31{,}202 & 533 \\
IBS pseudonym-grouped (B3)            & 176{,}943 & 58{,}945 & 25{,}776 & 24{,}228 & 24{,}634 & 662 \\
\bottomrule
\end{tabular}
\end{table}

\begin{figure}[t]
\centering
\includegraphics[width=\linewidth]{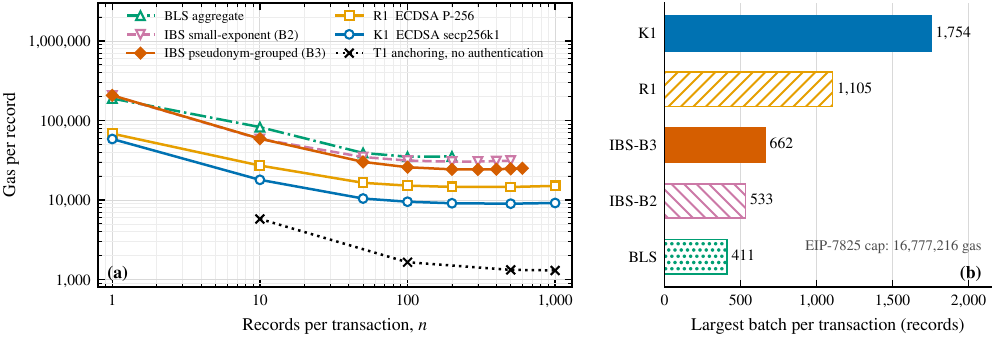}
\caption{(a) Gas per record versus records per transaction (log--log, $m=\min(n,10)$ devices); curves end where the next measured batch size exceeds the per-transaction cap. (b) Largest batch per transaction under the EIP-7825 cap, found by bisection with actual transactions.}
\label{fig:gas}
\end{figure}

\Cref{tab:gas} and \cref{fig:gas} summarize the measurements.

\paragraph{Individual submission.} Submitting each record in its own transaction costs between 55,475 (K1) and 187,118 gas (BLS). For ECDSA the cost is dominated by the 21,000 intrinsic gas, the 22,100-gas storage write of the record hash and the update of the device's timestamp, while signature verification itself costs only 3,000 (K1) or 6,900 gas (R1). For the pairing-based designs, the pairing check (102,900 gas for two pairs) dominates.

\paragraph{Batching.} Batching amortizes the intrinsic cost, the storage write and, for pairing-based designs, the pairing check. Per-record cost falls steeply up to about 100 records and is flat beyond (\cref{fig:gas}a); for very large batches it rises slightly because EVM memory expansion is priced quadratically. At $n=500$, K1 costs 8,963 gas per record, 6.2 times less than individual submission; R1 costs 14,542 gas (1.6 times K1). MLAV-IBS with pseudonym grouping costs 24,634 gas (2.7 times K1), 21\% less than the ungrouped small-exponent verifier. BLS aggregation costs 35,230 gas at $n=200$ (3.9 times K1), although it puts only one signature per batch in calldata.

\paragraph{Where the gas goes.} For K1, a batched record consists of about 224 bytes of calldata, one \texttt{ecrecover} (3,000 gas), two keccak256 hashes and a warm read and update of the device's timestamp. For BLS, hashing a message to $\G1$ costs 30,869 gas in our implementation. Its precompile floor is about 13,000 gas (two \texttt{MAP\_FP\_TO\_G1} calls at 5,500 gas, two MODEXP calls at the post-Fusaka minimum of 500 gas, one G1 addition and five SHA-256 calls), the rest being memory and encoding overhead. Even an implementation reaching this floor would cost more than 15,000 gas per record and remain above K1. For MLAV-IBS, each record contributes two points to G1 MSMs in B3 (three in B2), about 12,500 gas at the maximum MSM discount, plus about 420 bytes of calldata for two uncompressed EIP-2537 points and the record.

\paragraph{Fixed costs.} Deployment costs 483,408 (K1), 683,081 (R1), 942,740 (BLS) and 1,195,856 gas (IBS); registering a device costs 44,097--69,751 gas, including the initialisation of its timestamp slot.

\subsection{The transaction cap and block capacity}
\label{sec:capacity}

Since Fusaka, a transaction cannot exceed $2^{24}$ gas. The largest batch is 1,754 records for K1, 1,105 for R1, 662 for IBS-B3, 533 for IBS-B2 and 411 for BLS (\cref{fig:gas}b). Larger batches must be split; each additional transaction adds the fixed batch overhead of about 45,000--200,000 gas depending on the design, which is small relative to the batch. With the default block gas limit of 60 million~\cite{eip7935}, and ignoring other traffic, one block could anchor about 6,700 K1-authenticated records, 4,100 R1, 2,500 IBS-B3, 1,700 BLS or 45,000 records at trust level T1.

\subsection{Device and aggregator costs}

\begin{table}[t]
\centering
\caption{Off-chain costs (single core). Verification is per record for a batch of $n$ records from 10 devices; IBS uses pseudonym-grouped batching (B3), with $H_1$ evaluated once per pseudonym. Signature sizes are compressed encodings.}
\label{tab:offchain}
\small
\begin{tabular}{@{}lrrrrr@{}}
\toprule
& Sign (ms) & Signature & \multicolumn{3}{c}{Verify, ms per record} \\
\cmidrule(l){4-6}
Design & & bytes & $n=1$ & $n=100$ & $n=1000$ \\
\midrule
K1 ECDSA secp256k1 & 0.024 & 65 & 0.038 & 0.040 & 0.036 \\
R1 ECDSA P-256     & 0.026 & 64 & 0.080 & 0.087 & 0.072 \\
BLS aggregate      & 0.400 & 48 per batch & 1.771 & 0.316 & 0.256 \\
IBS grouped (B3)   & 0.398 & 96 & 2.221 & 0.093 & 0.038 \\
\bottomrule
\end{tabular}
\end{table}

\begin{table}[t]
\centering
\caption{Off-chain timings on five machines: the reference machine of \cref{tab:offchain} (Intel Xeon at 2.10\,GHz, one vCPU), GitHub-hosted runners with an ARM64 Arm Neoverse-N2 and an x86-64 AMD EPYC 9V74 (four vCPUs each), a Raspberry Pi 4 Model B (four Arm Cortex-A72 cores at 1.8\,GHz, 2\,GB RAM) and a Raspberry Pi Zero 2 W (four Arm Cortex-A53 cores at 1\,GHz, 512\,MB RAM), the class of board used in the IoP sensors. Single-threaded medians with the same package versions; the Raspberry Pi columns are medians of 10 runs each, whose coefficients of variation are at most 2.3\%; neither board throttled during any run.}
\label{tab:platforms}
\small\setlength{\tabcolsep}{3.5pt}
\begin{tabular}{@{}lrrrrr@{}}
\toprule
 & Xeon & Neoverse-N2 & EPYC 9V74 & Pi 4 & Pi Zero 2 W \\
 & (reference) & (ARM64) & (x86-64) & (Cortex-A72) & (Cortex-A53) \\
\midrule
\multicolumn{6}{@{}l}{\emph{Device signing (ms per record)}} \\
ECDSA secp256k1 (K1) & 0.024 & 0.028 & 0.026 & 0.128 & 0.287 \\
ECDSA P-256 (R1) & 0.026 & 0.025 & 0.024 & 0.128 & 0.344 \\
BLS & 0.400 & 0.433 & 0.399 & 2.384 & 5.699 \\
MLAV-IBS & 0.398 & 0.434 & 0.401 & 2.293 & 5.497 \\
\midrule
\multicolumn{6}{@{}l}{\emph{Aggregator verification (ms per record, $n=1{,}000$)}} \\
ECDSA secp256k1 (K1) & 0.035 & 0.045 & 0.047 & 0.258 & 0.582 \\
ECDSA P-256 (R1) & 0.072 & 0.067 & 0.064 & 0.321 & 0.799 \\
BLS & 0.256 & 0.258 & 0.259 & 1.306 & 3.028 \\
MLAV-IBS & 0.038 & 0.035 & 0.036 & 0.170 & 0.403 \\
MLAV-IBS, $Q$ cached & 0.031 & 0.033 & 0.034 & 0.157 & 0.373 \\
\bottomrule
\end{tabular}
\end{table}

\begin{figure}[t]
\centering
\includegraphics[width=84mm]{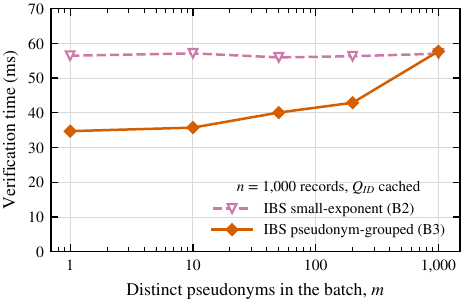}
\caption{Off-chain MLAV-IBS batch verification time for 1,000 records versus the number of distinct pseudonyms ($Q_{\ID}$ cached). Grouping helps most when few pseudonyms produce many records.}
\label{fig:devices}
\end{figure}

\Cref{tab:offchain} reports off-chain costs. Pairing-based signing is about 16 times slower than ECDSA signing on the reference machine (0.40 versus 0.024\,ms), about 19 times slower on a Raspberry Pi 4 and about 20 times slower on the Raspberry Pi Zero 2 W of the IoP sensor (5.7 versus 0.29\,ms, \cref{tab:platforms}). The relative gap is large, but in absolute terms a few milliseconds per record are negligible for a sensor that signs one record every few minutes, so for battery-powered devices how often they sign and transmit matters more than which of these schemes they use. For the aggregator, ECDSA verification costs 0.036\,ms (K1) and 0.072\,ms (R1) per record at any batch size. Batch verification lets MLAV-IBS match K1 for large batches (0.038\,ms per record at $n=1000$), whereas BLS remains 7 times slower than K1 because every message must be hashed to the curve. Pseudonym grouping matters most when $H_1$ must be evaluated: for 1,000 records from 50 pseudonyms, B3 takes 64\,ms against 351\,ms for B2; with $H_1$ cached, the saving is 35--40\% for few pseudonyms and vanishes when every record comes from a different pseudonym (\cref{fig:devices}).

\paragraph{Aggregator throughput on several cores.} An aggregator verifies independent batches in parallel, so its throughput scales with the number of physical cores (\cref{tab:parallel}): with four worker processes on four cores, every design verifies 3.8 to 4.0 times as many records per second as with one, on the Neoverse-N2 server cores as on both Raspberry Pi boards; only R1 on the Pi Zero 2 W scales less (3.54 times, consistently across 10 runs), whereas on the Xeon runner the third and fourth workers share two cores with the first two and hardware threads add only 19--26\%. Even one core is far from being the bottleneck. At about 6,700 K1-authenticated records per 12-second block (\cref{sec:capacity}), the chain accepts about 560 records per second, whereas one Neoverse-N2 core verifies about 22,000; for BLS, about 140 records per second against about 3,900, and for IBS-B3 about 210 against about 31,000. Even a Raspberry Pi Zero 2 W with four workers verifies 6,812 K1, 1,320 BLS and 10,471 IBS records per second, 9 to 50 times what the chain can accept. The cost of authenticating IoT records is therefore set by the chain's gas, not by the aggregator's computation, and even a low-cost board can serve the full capacity of the chain.

\begin{table}[t]
\centering
\caption{Aggregator throughput: records verified per second with 1, 2 and 4 worker processes, each verifying pre-signed batches of 1,000 records from 10 devices, and the speedup over one worker. The Xeon runner has two physical cores with two hardware threads each; the other machines have four cores with one thread each. Raspberry Pi rows are medians of 10 runs, whose coefficients of variation are at most 1.1\%; neither board throttled.}
\label{tab:parallel}
\footnotesize\setlength{\tabcolsep}{3pt}
\begin{tabular}{@{}lrrrr@{}}
\toprule
Machine, workers (w.) & K1 & R1 & BLS & IBS ($Q$ cached) \\
\midrule
Neoverse-N2 (4 cores), 1 w. & 22{,}047 & 14{,}889 & 3{,}874 & 30{,}920 \\
\quad 4 w. & 87{,}464 (3.97$\times$) & 59{,}098 (3.97$\times$) & 15{,}391 (3.97$\times$) & 123{,}347 (3.99$\times$) \\
\midrule
Xeon 6975P-C (2 cores $\times$ 2 threads), 1 w. & 28{,}884 & 15{,}543 & 4{,}496 & 34{,}627 \\
\quad 2 w. & 56{,}927 (1.97$\times$) & 29{,}717 (1.91$\times$) & 8{,}837 (1.97$\times$) & 69{,}271 (2.00$\times$) \\
\quad 4 w. & 68{,}025 (2.36$\times$) & 35{,}435 (2.28$\times$) & 11{,}108 (2.47$\times$) & 87{,}604 (2.53$\times$) \\
\midrule
Raspberry Pi 4 (4 cores), 1 w. & 3{,}884 & 3{,}123 & 766 & 6{,}376 \\
\quad 4 w. & 15{,}158 (3.90$\times$) & 12{,}075 (3.86$\times$) & 3{,}023 (3.95$\times$) & 24{,}439 (3.83$\times$) \\
\midrule
Raspberry Pi Zero 2 W (4 cores), 1 w. & 1{,}718 & 1{,}250 & 330 & 2{,}683 \\
\quad 4 w. & 6{,}812 (3.96$\times$) & 4{,}421 (3.54$\times$) & 1{,}320 (4.00$\times$) & 10{,}471 (3.90$\times$) \\
\bottomrule
\end{tabular}
\end{table}

\subsection{Before and after Pectra and Fusaka}
\label{sec:forks}

\begin{table}[t]
\centering
\caption{The same transactions under the rules before Pectra (Cancun), after Pectra (Prague) and after Fusaka (Osaka): gas per record, $m=\min(n,10)$ devices. ``n/a'': the precompile the design needs is not available.}
\label{tab:forks}
\small\setlength{\tabcolsep}{4pt}
\begin{tabular}{@{}lrrr@{}}
\toprule
Design & Cancun & Prague & Osaka \\
\midrule
T1 record hashes, $n=500$ & 1{,}286 & 1{,}320 & 1{,}319 \\
K1 ECDSA secp256k1, $n=500$ & 8{,}964 & 8{,}963 & 8{,}963 \\
R1 ECDSA P-256, $n=500$ & n/a & n/a & 14{,}542 \\
BLS aggregate, $n=100$ & n/a & 34{,}328 & 34{,}927 \\
BLS aggregate, $n=500$ & n/a & 42{,}251 & above cap \\
IBS pseudonym-grouped (B3), $n=500$ & n/a & 24{,}633 & 24{,}634 \\
\midrule
Largest transaction & 36\,M (block) & 36\,M (block) & $2^{24}$ (EIP-7825) \\
\bottomrule
\end{tabular}
\end{table}

To isolate what the two upgrades changed, we executed the same contracts and transactions under the Cancun rules that preceded Pectra and the Prague rules of Pectra (\cref{tab:forks}). Their main effect is on feasibility rather than cost. Before Pectra, only K1 can be verified with a precompile; Pectra's BLS12-381 precompiles (EIP-2537) make BLS and MLAV-IBS possible, and Fusaka's P-256 precompile (EIP-7951) adds R1. Designs available under several rules cost almost the same under each: K1 within 0.01\%, and IBS within a few gas. Two repricings are visible. Fusaka's MODEXP repricing (EIP-7883) adds 600 gas to every hash to $\G1$ (30,269 against 30,869 gas), about 1.7\% per BLS record at $n=100$. Pectra's calldata floor (EIP-7623) raises T1, which posts data with little execution, by 2.6\%, and binds for none of the authenticated designs. The largest change is the per-transaction cap of Fusaka. Before it, a transaction was bounded only by the block gas limit, 36 million gas when Pectra activated, so a BLS batch of 500 records (42,251 gas per record) and 1,000 per-slot writes (22.8 million gas) were possible; under Osaka both exceed $2^{24}$ gas. The cap therefore matters for designs whose cost grows with the batch, not for the cheapest ones.

\subsection{Alternatives we did not implement}
\label{sec:alternatives}

\paragraph{Record hashes in a blob (T1).} A blob carries 4,096 field elements of which 31 bytes each are usable, enough for 3,968 record hashes~\cite{eip4844}. Its gas is priced in a separate market whose base fee is, since Fusaka, bounded below so that a blob costs at least as much as 8,192 gas of execution~\cite{eip7918}. Posting the hashes of a T1 batch in a blob, and only the root and the blob's versioned hash in the transaction, would cost about one T0 transaction with a second storage write plus one blob, about 76,000 gas for 3,968 records, i.e.\ roughly 20 gas per record at the reserve price against 1,319 for T1 (an estimate; we did not execute blob transactions). The saving has two costs. A contract can read only a blob's commitment, not its contents, so it can no longer compute the root itself and must accept the aggregator's claim that the root matches the blob, unless each record is opened with the point-evaluation precompile. And blob contents are kept for only about 18 days, after which the publicity that defines T1 depends on archivers. PeerDAS~\cite{eip7594}, introduced with Fusaka, lets the network carry more blobs, which makes this option more available but not more trustworthy. A blob is therefore a cheaper data-availability layer for T0 rather than a substitute for T1.

\paragraph{Proof-verified aggregation (T2).} Instead of verifying every signature on-chain, an aggregator can verify the signatures and the replay rule off-chain and prove in a succinct argument that the posted root commits only to records signed by registered devices with fresh timestamps. A Groth16 proof~\cite{groth2016size} over BN254 is verified with one four-pair pairing check, 181,000 gas under EIP-1108~\cite{eip1108}, plus a 6,000-gas scalar multiplication per public input. With the root and the old and new commitments to the device timestamps as public inputs, the 21,000-gas base, a 256-byte proof and two storage writes, a batch would cost about 260,000 gas regardless of its size (an estimate): less than K1 from about 30 records per batch, and about 260 gas per record at $n=1{,}000$. The per-device timestamps must move into the committed state that the proof updates; otherwise the per-device storage updates of \cref{sec:space} remain. The costs move off-chain. secp256k1 signatures must be verified with non-native arithmetic inside the circuit, which makes proving far more expensive than verifying the signatures directly and adds the proving time to the anchoring latency; Groth16 also needs a trusted setup per circuit, and pairing-based proofs are not post-quantum. Proof-verified aggregation is attractive for large batches when an aggregator with proving hardware is available; implementing it and measuring the prover are left to future work.

\subsection{Qualitative properties}

\begin{table}[t]
\centering
\caption{Properties of the designs. The KGC registers each device once; in K1, R1 and BLS the device generates its own key, whereas in IBS the KGC derives it.}
\label{tab:properties}
\small\setlength{\tabcolsep}{3.5pt}
\begin{tabular}{@{}lcccc@{}}
\toprule
Property & K1 & R1 & BLS & IBS \\
\midrule
Supported by device secure elements & partial & yes & no & no \\
Key material per device on-chain & address & commitment & commitment & commitment \\
Public-key distribution needed off-chain & yes & yes & yes & no \\
Signatures in calldata & per record & per record & one per batch & per record \\
Pseudonymity with KGC tracing & via KGC records & via KGC records & via KGC records & built in \\
Device compromise exposes & own key & own key & own key & own key \\
KGC can sign as a registered device & no & no & no & yes (escrow) \\
\bottomrule
\end{tabular}
\end{table}

\Cref{tab:properties} compares properties that do not show up in gas. All designs can provide pseudonymity towards outsiders with tracing by the KGC, ECDSA and BLS by having the KGC issue and record per-device keys; the identity-based design makes this mapping cryptographic and removes the need to distribute public keys off-chain, an advantage that the on-chain registry largely neutralizes.

\paragraph{Key escrow.} The identity-based design pays for this convenience with key escrow. The KGC computes every device key, $\sk_{\ID}=s\,H_1(\ID)$, and can therefore sign records in the name of any device, and such signatures are indistinguishable from the device's own. Since the consortium that operates the KGC includes the lender, the lender could fabricate the production records that release its own payments, which is exactly what on-chain authentication (T2) is meant to rule out. In the other designs the device generates its key and the KGC only registers it; because the registry accepts each device once, the KGC can add devices but cannot sign as an existing one or replace its key, and every registration is a public transaction. The escrow can be removed at additional cost by splitting the master key among consortium members with threshold key extraction~\cite{boneh2001ibe} or by certificateless signatures, in which part of the key is chosen by the device~\cite{alriyami2003certificateless}; we did not implement either. Built-in pseudonymity with conditional traceability is therefore the only advantage of the identity-based design, and it comes at the price of trusting the KGC with the authenticity of every record.

\section{Findings and guidelines}
\label{sec:guidelines}

\begin{finding}
Anchoring is the first lever. Merkle anchoring reduces the cost of recording a hash from 22,780 gas (one storage slot per record) to 1,319 gas when all hashes are published (T1), and to a constant 45,507 gas per batch when only the root is posted (T0). Any design that writes one storage slot per record should be avoided.
\end{finding}

\begin{finding}
On-chain authentication costs a factor of six to seven. Moving from T1 to T2 with the cheapest scheme raises the cost from 1,319 to 8,963 gas per record. Whether this is worthwhile depends on whether the aggregator can be trusted; in a consortium in which the lender itself operates the aggregator, T1 may be sufficient, whereas T2 is needed when records serve as evidence towards parties outside the consortium.
\end{finding}

\begin{finding}
For on-chain authentication, ECDSA over secp256k1 with batching is the cheapest design measured, at 8,963 gas per record and up to 1,754 records per transaction. P-256 costs 1.6 times as much but matches the hardware that IoT devices typically embed. Pairing-based designs cost 2.7--3.9 times as much on-chain and are 16 times slower to sign with.
\end{finding}

\begin{finding}
BLS aggregation does not pay off on-chain for IoT records. Aggregation removes per-record signatures from calldata, but every message must still be hashed to the curve on-chain, which costs more than verifying an ECDSA signature. It also yields the smallest batches under the transaction cap (411 records).
\end{finding}

\begin{finding}
Pairing-based identity-based signatures are justified by properties, not by cost. Their batch verification makes them competitive off-chain with ECDSA for large batches, and pseudonym grouping reduces their on-chain cost by 22\% and their off-chain cost by up to 5.5 times when identities must be hashed. They are worth their premium when built-in pseudonymity with conditional traceability is a requirement, and only if the scheme is chosen with care: signatures of the IBV family, as used in the original MLAV prototype, are insecure.
\end{finding}

\paragraph{Converting gas to fees.} The fee of a record is its gas multiplied by the prevailing gas price; for example, at a base fee of $g$ gwei and an ether price of $E$ USD, a K1 record in a batch costs $8{,}963\cdot g\cdot E\cdot10^{-9}$ USD. Because the ranking of designs depends only on gas, our conclusions hold for any fee level. On rollups that execute the same precompiles, execution gas is the same, while the cost of calldata depends on the rollup's data-availability pricing; designs with less calldata per record, such as BLS aggregation, would gain relative to ECDSA there.

\section{Limitations}
\label{sec:limitations}

Our implementations are straightforward Solidity without hand-written assembly beyond buffer handling; optimized code, in particular for BLS hash-to-curve, would lower the absolute costs of the pairing-based designs, but our precompile-floor analysis indicates that it would not change the ranking. Off-chain timings were taken on server cores and on two Raspberry Pi boards of the class used in the IoP sensors, not on sensors deployed in the field, and we did not measure energy. The blob-based and proof-verified alternatives of \cref{sec:alternatives} are estimated from protocol prices, not implemented. The workload is synthetic, with 10 devices per batch; we have not evaluated localization of invalid signatures in a failed batch, which requires bisection~\cite{ferrara2009practical}, nor deployment on rollups. Finally, gas schedules change with network upgrades; our harness makes it straightforward to repeat the study under future rules, as we did for the rules that preceded Pectra and Fusaka (\cref{sec:forks}).

\section{Conclusion}
\label{sec:conclusion}

We measured what it costs to authenticate and anchor IoT records on Ethereum after the Pectra and Fusaka upgrades. Merkle anchoring and batching are the decisive levers, reducing the cost per record by more than an order of magnitude. Among authenticated designs, batched secp256k1 ECDSA is the cheapest at 8,963 gas per record, while the new BLS12-381 precompiles make pairing-based designs feasible but not competitive in cost: they are justified only by additional properties such as built-in pseudonymity with conditional traceability, which the identity-based design obtains at the price of key escrow at the KGC. For agricultural supply chain finance, this suggests anchoring record hashes at trust level T1 within a lender-operated consortium and adding batched ECDSA or P-256 authentication when records must convince outside parties. Future work includes energy measurements on deployed sensors, an implementation of proof-verified aggregation with a measurement of its prover (\cref{sec:alternatives}), deployment on rollups and an assessment of privacy-preserving credit evidence based on zero-knowledge proofs.

\section*{Acknowledgements}
The MLAV architecture, the camera prototype and the supply chain finance workflow on which this study builds were developed with Jingze Wu, Seth Austin Harding, Chien-Lung Lin, San-Tai Wang and Bo-Yen Tang at National Taiwan University~\cite{wu2021mlav,sie2022secured}; the authors gratefully acknowledge their contributions. The camera prototype and the first smart-contract mock-up of the platform originated in the final project ``Agricultural Blockchain: A Reconceptualized Approach to Organic'' of the 2019 blockchain course at National Taiwan University; we thank its student team.

\section*{Declarations}

\paragraph{Funding.} This research did not receive any specific grant from funding agencies in the public, commercial, or not-for-profit sectors.

\paragraph{Competing interests.} Ming-Fong Sie was a co-founder and the chief executive officer, and Shih-wei Liao a co-founder, of Conzen.AI, a company that developed privacy-preserving computation for financial compliance and supply chain traceability and was dissolved in October 2026. The company did not fund this study and had no role in its design, analysis or reporting. The other authors declare no competing interests.

\paragraph{Ethics approval and consent to participate.} Not applicable.

\paragraph{Consent for publication.} Not applicable.

\paragraph{Data and code availability.} The contracts, benchmark harnesses, scripts reproducing the attacks on the original MLAV signature, and all raw measurements are available at \url{https://github.com/Siemingfong/iot-auth-cost-study} (release v1.1), archived at Zenodo under DOI \href{https://doi.org/10.5281/zenodo.24328819}{10.5281/zenodo.24328819}.

\paragraph{Author contributions.} \textbf{Ming-Fong Sie}: conceptualization, methodology, software, investigation, formal analysis, data curation, visualization, writing -- original draft. \textbf{Jian-Hong Chen}: investigation, software, validation, writing -- review \& editing. \textbf{Gwan-yen Lin}: investigation, validation, writing -- review \& editing. \textbf{Yen-Shuo Chen}: formal analysis, validation, writing -- review \& editing. \textbf{Shih-wei Liao}: conceptualization, supervision, resources, writing -- review \& editing.

\paragraph{Use of generative AI.} The authors used Claude (Anthropic) to assist with writing the benchmark code and Solidity contracts, running the experiments, analysing data, preparing figures, checking the security argument of the appendix, and drafting and editing the text. All code, measurements and text were reviewed by the authors, who take full responsibility for the content of the publication.

\appendix
\section{Unforgeability of MLAV-IBS (proof sketch)}
\label{app:proof}

\paragraph{Model and assumption.} We use existential unforgeability under adaptive chosen-message and chosen-identity attacks (EUF-ID-CMA): the adversary receives the public parameters, may adaptively query $H_1$, $H_2$, a key-extraction oracle and a signing oracle, and wins by outputting a valid signature on $(\ID^*,M^*)$ for an $\ID^*$ it never extracted and a pair $(\ID^*,M^*)$ it never submitted to the signing oracle. The reduction targets the following computational co-Diffie--Hellman problem in the Type-3 setting~\cite{galbraith2008pairings}: given $(P_1,aP_1,bP_1,P_2,aP_2)$ for uniform $a,b\in\Zr$, compute $abP_1$. The element $aP_2$, which the plain computational Diffie--Hellman problem in $\G1$ lacks, is needed to publish $\Ppub{}_{,2}$.

\paragraph{Reduction.} The argument follows Cha and Cheon~\cite{cha2003ibs} in the random oracle model for $H_1$ and $H_2$. $H_1$ is instantiated with the random-oracle variant of the RFC~9380 hash-to-curve, which is designed to be indifferentiable from a random oracle~\cite{rfc9380}. The simulator sets $\Ppub{}_{,1}=aP_1$ and $\Ppub{}_{,2}=aP_2$, guesses the target pseudonym $\ID^*$ among the $q_{H_1}$ queries to $H_1$, programs $H_1(\ID^*)=bP_1$, and answers other $H_1$ queries with $x_{\ID}P_1$ for uniform $x_{\ID}$. Extraction queries for $\ID\neq\ID^*$ are answered with $\sk_{\ID}=x_{\ID}\,\Ppub{}_{,1}$; an extraction query for $\ID^*$ means the guess was wrong, and the simulator aborts. Signing queries for any pseudonym are answered without the key: the simulator samples $h$ with the distribution of $H_2$'s output (for keccak256 reduced modulo $r$, which is not exactly uniform on $\Zr$, by reducing a uniform 256-bit string), samples $y\xleftarrow{\$}\Zr$, sets $U=yP_1-hQ_{\ID}$ and $V=y\,\Ppub{}_{,1}$, and programs $H_2(\ID,M,U)=h$; indeed $e(V,P_2)=e(yP_1,aP_2)=e(U+hQ_{\ID},\Ppub{}_{,2})$. Since $U$ is uniform in $\G1$, the programmed point is already defined only with probability at most $(q_{H_2}+q_S)\,q_S/r$, in which case the simulator aborts. From a forgery $(U,h,V)$ on $\ID^*$, the forking lemma~\cite{pointcheval2000security} yields a second forgery with the same $U$ and a different $h'$. Every valid signature satisfies $e(V,P_2)=e(U+hQ_{\ID^*},aP_2)$, and since the pairing is non-degenerate this means $V=a(U+hQ_{\ID^*})$, whatever way the forger produced it; hence $V-V'=(h-h')\,aQ_{\ID^*}$ and $(h-h')^{-1}(V-V')=abP_1$. Since $H_2$'s outputs lie in $\Zr$ and collide with negligible probability, $h-h'$ is invertible modulo $r$.

\paragraph{Tightness and batches.} The reduction loses a factor of about $q_{H_1}$ for guessing $\ID^*$ and, through the forking lemma, roughly squares the adversary's advantage, so it is not tight, as in Cha and Cheon's analysis. Combined with \cref{thm:batch}, an accepted batch contains a forged record only if the adversary breaks the unforgeability of a single signature or the batch test, so accepted batches are unforgeable up to the sum of both bounds, the bound of \cref{thm:batch} counted once per submitted batch; on-chain, $H_2$ additionally binds the domain $d$, which prevents reuse of signatures across chains and contracts. 

\bibliographystyle{unsrtnat}
\bibliography{references}

\begin{thebibliography}{57}
\providecommand{\natexlab}[1]{#1}
\providecommand{\url}[1]{\texttt{#1}}
\expandafter\ifx\csname urlstyle\endcsname\relax
  \providecommand{\doi}[1]{doi: #1}\else
  \providecommand{\doi}{doi: \begingroup \urlstyle{rm}\Url}\fi

\bibitem[Tripoli and Schmidhuber(2018)]{tripoli2018emerging}
Mischa Tripoli and Josef Schmidhuber.
\newblock Emerging opportunities for the application of blockchain in the
  agri-food industry.
\newblock \emph{FAO and ICTSD: Rome and Geneva}, 2018.

\bibitem[Kamilaris et~al.(2019)Kamilaris, Fonts, and
  Prenafeta-Bold{\'u}]{kamilaris2019rise}
Andreas Kamilaris, Agusti Fonts, and Francesc~X. Prenafeta-Bold{\'u}.
\newblock The rise of blockchain technology in agriculture and food supply
  chains.
\newblock \emph{Trends in Food Science \& Technology}, 91:\penalty0 640--652,
  2019.

\bibitem[Wang et~al.(2019)Wang, Han, and Beynon-Davies]{wang2019understanding}
Yingli Wang, Jeong~Hugh Han, and Paul Beynon-Davies.
\newblock Understanding blockchain technology for future supply chains: a
  systematic literature review and research agenda.
\newblock \emph{Supply Chain Management: An International Journal}, 24\penalty0
  (1):\penalty0 62--84, 2019.

\bibitem[Caro et~al.(2018)Caro, Ali, Vecchio, and
  Giaffreda]{caro2018blockchain}
Miguel~Pincheira Caro, Muhammad~Salek Ali, Massimo Vecchio, and Raffaele
  Giaffreda.
\newblock Blockchain-based traceability in agri-food supply chain management: A
  practical implementation.
\newblock In \emph{2018 IoT Vertical and Topical Summit on Agriculture --
  Tuscany (IOT Tuscany)}, pages 1--4, Piscataway, NJ, USA, 2018. IEEE.

\bibitem[Bhat et~al.(2022)Bhat, Huang, Sofi, and Sultan]{bhat2022agri}
Shabir~Ahmad Bhat, Nen-Fu Huang, Ishfaq~Bashir Sofi, and Muhammad Sultan.
\newblock Agriculture-food supply chain management based on blockchain and
  {IoT}: A narrative on enterprise blockchain interoperability.
\newblock \emph{Agriculture}, 12\penalty0 (1):\penalty0 40, 2022.

\bibitem[Chod et~al.(2020)Chod, Trichakis, Tsoukalas, Aspegren, and
  Weber]{chod2020financing}
Jiri Chod, Nikolaos Trichakis, Gerry Tsoukalas, Henry Aspegren, and Mark Weber.
\newblock On the financing benefits of supply chain transparency and blockchain
  adoption.
\newblock \emph{Management Science}, 66\penalty0 (10):\penalty0 4378--4396,
  2020.

\bibitem[eip(2025{\natexlab{a}})]{eip7600}
{EIP-7600}: Hardfork meta -- pectra.
\newblock Ethereum Improvement Proposals,
  \url{https://eips.ethereum.org/EIPS/eip-7600}, 2025{\natexlab{a}}.

\bibitem[eip(2020)]{eip2537}
{EIP-2537}: Precompile for {BLS12-381} curve operations.
\newblock Ethereum Improvement Proposals,
  \url{https://eips.ethereum.org/EIPS/eip-2537}, 2020.
\newblock Activated on mainnet in the Pectra upgrade (2025).

\bibitem[eip(2025{\natexlab{b}})]{eip7623}
{EIP-7623}: Increase calldata cost.
\newblock Ethereum Improvement Proposals,
  \url{https://eips.ethereum.org/EIPS/eip-7623}, 2025{\natexlab{b}}.

\bibitem[eip(2025{\natexlab{c}})]{eip7607}
{EIP-7607}: Hardfork meta -- fusaka.
\newblock Ethereum Improvement Proposals,
  \url{https://eips.ethereum.org/EIPS/eip-7607}, 2025{\natexlab{c}}.

\bibitem[{Ethereum Foundation Protocol Coordination Team}(2025)]{ef2025fusaka}
{Ethereum Foundation Protocol Coordination Team}.
\newblock Fusaka mainnet announcement.
\newblock
  \url{https://blog.ethereum.org/2025/11/06/fusaka-mainnet-announcement}, 2025.

\bibitem[eip(2025{\natexlab{d}})]{eip7951}
{EIP-7951}: Precompile for secp256r1 curve support.
\newblock Ethereum Improvement Proposals,
  \url{https://eips.ethereum.org/EIPS/eip-7951}, 2025{\natexlab{d}}.

\bibitem[eip(2025{\natexlab{e}})]{eip7883}
{EIP-7883}: Modexp gas cost increase.
\newblock Ethereum Improvement Proposals,
  \url{https://eips.ethereum.org/EIPS/eip-7883}, 2025{\natexlab{e}}.

\bibitem[eip(2025{\natexlab{f}})]{eip7935}
{EIP-7935}: Set default gas limit to 60m.
\newblock Ethereum Improvement Proposals,
  \url{https://eips.ethereum.org/EIPS/eip-7935}, 2025{\natexlab{f}}.

\bibitem[eip(2025{\natexlab{g}})]{eip7825}
{EIP-7825}: Transaction gas limit cap.
\newblock Ethereum Improvement Proposals,
  \url{https://eips.ethereum.org/EIPS/eip-7825}, 2025{\natexlab{g}}.

\bibitem[Wu et~al.(2021)Wu, Sie, Harding, Lin, Wang, and Liao]{wu2021mlav}
Jingze Wu, Ming-Fong Sie, Seth~Austin Harding, Chien-Lung Lin, San-Tai Wang,
  and Shih-wei Liao.
\newblock Multi-layer aggregate verification for {IoT} blockchain.
\newblock In \emph{2021 3rd Conference on Blockchain Research \& Applications
  for Innovative Networks and Services (BRAINS)}, pages 43--44, Piscataway, NJ,
  USA, 2021. IEEE.
\newblock \doi{10.1109/BRAINS52497.2021.9569817}.

\bibitem[Sie et~al.(2022)Sie, Wu, Harding, Lin, Wang, and Liao]{sie2022secured}
Ming-Fong Sie, Jingze Wu, Seth~Austin Harding, Chien-Lung Lin, San-Tai Wang,
  and Shih-wei Liao.
\newblock Secured multi-layer blockchain framework for {IoT} aggregate
  verification.
\newblock \emph{Advances in Science, Technology and Engineering Systems
  Journal}, 7\penalty0 (3):\penalty0 106--115, 2022.
\newblock \doi{10.25046/aj070312}.

\bibitem[Zhang et~al.(2008)Zhang, Lu, Lin, Ho, and Shen]{zhang2008ibv}
Chenxi Zhang, Rongxing Lu, Xiaodong Lin, Pin-Han Ho, and Xuemin Shen.
\newblock An efficient identity-based batch verification scheme for vehicular
  sensor networks.
\newblock In \emph{IEEE INFOCOM 2008}, pages 246--250, 2008.

\bibitem[Lee and Lai(2013)]{lee2013toward}
Cheng-Chi Lee and Yan-Ming Lai.
\newblock Toward a secure batch verification with group testing for {VANET}.
\newblock \emph{Wireless Networks}, 19\penalty0 (6):\penalty0 1441--1449, 2013.

\bibitem[Tian(2016)]{tian2016agri}
Feng Tian.
\newblock An agri-food supply chain traceability system for {China} based on
  {RFID} \& blockchain technology.
\newblock In \emph{2016 13th International Conference on Service Systems and
  Service Management (ICSSSM)}, pages 1--6, Piscataway, NJ, USA, 2016. IEEE.

\bibitem[Tian(2017)]{tian2017supply}
Feng Tian.
\newblock A supply chain traceability system for food safety based on {HACCP},
  blockchain \& {Internet of Things}.
\newblock In \emph{2017 International Conference on Service Systems and Service
  Management}, pages 1--6, Piscataway, NJ, USA, 2017. IEEE.

\bibitem[Tseng et~al.(2018)Tseng, Liao, Chong, and Liao]{tseng2018governance}
Jen-Hung Tseng, Yen-Chih Liao, Bin Chong, and Shih-wei Liao.
\newblock Governance on the drug supply chain via {Gcoin} blockchain.
\newblock \emph{International Journal of Environmental Research and Public
  Health}, 15\penalty0 (6):\penalty0 1055, 2018.

\bibitem[Dorri et~al.(2017)Dorri, Kanhere, Jurdak, and
  Gauravaram]{dorri2017blockchain}
Ali Dorri, Salil~S. Kanhere, Raja Jurdak, and Praveen Gauravaram.
\newblock Blockchain for {IoT} security and privacy: The case study of a smart
  home.
\newblock In \emph{2017 IEEE International Conference on Pervasive Computing
  and Communications Workshops (PerCom Workshops)}, pages 618--623, 2017.

\bibitem[Novo(2018)]{novo2018blockchain}
Oscar Novo.
\newblock Blockchain meets {IoT}: An architecture for scalable access
  management in {IoT}.
\newblock \emph{IEEE Internet of Things Journal}, 5\penalty0 (2):\penalty0
  1184--1195, 2018.

\bibitem[Lee and Lee(2017)]{lee2017blockchain}
Boohyung Lee and Jong-Hyouk Lee.
\newblock Blockchain-based secure firmware update for embedded devices in an
  {Internet of Things} environment.
\newblock \emph{The Journal of Supercomputing}, 73\penalty0 (3):\penalty0
  1152--1167, 2017.

\bibitem[Hu et~al.(2019)Hu, Yeh, Liao, and Yang]{hu2019autonomous}
Jen-Wei Hu, Lo-Yao Yeh, Shih-Wei Liao, and Chu-Sing Yang.
\newblock Autonomous and malware-proof blockchain-based firmware update
  platform with efficient batch verification for {Internet of Things} devices.
\newblock \emph{Computers \& Security}, 86:\penalty0 238--252, 2019.

\bibitem[Thibault et~al.(2022)Thibault, Sarry, and Hafid]{thibault2022rollups}
Louis~Tremblay Thibault, Tom Sarry, and Abdelhakim~Senhaji Hafid.
\newblock Blockchain scaling using rollups: A comprehensive survey.
\newblock \emph{IEEE Access}, 10:\penalty0 93039--93054, 2022.

\bibitem[Boneh et~al.(2001)Boneh, Lynn, and Shacham]{boneh2001short}
Dan Boneh, Ben Lynn, and Hovav Shacham.
\newblock Short signatures from the {Weil} pairing.
\newblock In \emph{Advances in Cryptology -- ASIACRYPT 2001}, volume 2248 of
  \emph{LNCS}, pages 514--532, Berlin, Heidelberg, 2001. Springer.

\bibitem[Bowe(2017)]{bowe2017bls12}
Sean Bowe.
\newblock {BLS12-381}: New {zk-SNARK} elliptic curve construction.
\newblock Electric Coin Company blog,
  \url{https://electriccoin.co/blog/new-snark-curve/}, 2017.

\bibitem[Boneh et~al.(2003)Boneh, Gentry, Lynn, and
  Shacham]{boneh2003aggregate}
Dan Boneh, Craig Gentry, Ben Lynn, and Hovav Shacham.
\newblock Aggregate and verifiably encrypted signatures from bilinear maps.
\newblock In \emph{Advances in Cryptology -- EUROCRYPT 2003}, volume 2656 of
  \emph{LNCS}, pages 416--432, Berlin, Heidelberg, 2003. Springer.

\bibitem[Boneh et~al.(2018)Boneh, Drijvers, and Neven]{boneh2018compact}
Dan Boneh, Manu Drijvers, and Gregory Neven.
\newblock Compact multi-signatures for smaller blockchains.
\newblock In \emph{Advances in Cryptology -- ASIACRYPT 2018}, volume 11273 of
  \emph{LNCS}, pages 435--464, Cham, 2018. Springer.

\bibitem[Faz-Hern{\'a}ndez et~al.(2023)Faz-Hern{\'a}ndez, Scott, Sullivan,
  Wahby, and Wood]{rfc9380}
Armando Faz-Hern{\'a}ndez, Sam Scott, Nick Sullivan, Riad~S. Wahby, and
  Christopher~A. Wood.
\newblock Hashing to elliptic curves.
\newblock RFC 9380, Internet Research Task Force, 2023.

\bibitem[Shamir(1985)]{shamir1984identity}
Adi Shamir.
\newblock Identity-based cryptosystems and signature schemes.
\newblock In \emph{Advances in Cryptology -- CRYPTO '84}, volume 196 of
  \emph{LNCS}, pages 47--53, Berlin, Heidelberg, 1985. Springer.

\bibitem[Boneh and Franklin(2001)]{boneh2001ibe}
Dan Boneh and Matt Franklin.
\newblock Identity-based encryption from the {Weil} pairing.
\newblock In \emph{Advances in Cryptology -- CRYPTO 2001}, volume 2139 of
  \emph{LNCS}, pages 213--229, Berlin, Heidelberg, 2001. Springer.

\bibitem[Cha and Cheon(2003)]{cha2003ibs}
Jae~Choon Cha and Jung~Hee Cheon.
\newblock An identity-based signature from gap {Diffie-Hellman} groups.
\newblock In \emph{Public Key Cryptography -- PKC 2003}, volume 2567 of
  \emph{LNCS}, pages 18--30, Berlin, Heidelberg, 2003. Springer.

\bibitem[Hess(2003)]{hess2002ibs}
Florian Hess.
\newblock Efficient identity based signature schemes based on pairings.
\newblock In \emph{Selected Areas in Cryptography -- SAC 2002}, volume 2595 of
  \emph{LNCS}, pages 310--324, Berlin, Heidelberg, 2003. Springer.

\bibitem[Bellare et~al.(1998)Bellare, Garay, and Rabin]{bellare1998batch}
Mihir Bellare, Juan~A. Garay, and Tal Rabin.
\newblock Fast batch verification for modular exponentiation and digital
  signatures.
\newblock In \emph{Advances in Cryptology -- EUROCRYPT '98}, volume 1403 of
  \emph{LNCS}, pages 236--250, Berlin, Heidelberg, 1998. Springer.

\bibitem[Camenisch et~al.(2007)Camenisch, Hohenberger, and
  Pedersen]{camenisch2007batch}
Jan Camenisch, Susan Hohenberger, and Michael~{\O}stergaard Pedersen.
\newblock Batch verification of short signatures.
\newblock In \emph{Advances in Cryptology -- EUROCRYPT 2007}, volume 4515 of
  \emph{LNCS}, pages 246--263, Berlin, Heidelberg, 2007. Springer.

\bibitem[Ferrara et~al.(2009)Ferrara, Green, Hohenberger, and
  Pedersen]{ferrara2009practical}
Anna~Lisa Ferrara, Matthew Green, Susan Hohenberger, and Michael~{\O}stergaard
  Pedersen.
\newblock Practical short signature batch verification.
\newblock In \emph{Topics in Cryptology -- CT-RSA 2009}, volume 5473 of
  \emph{LNCS}, pages 309--324, Berlin, Heidelberg, 2009. Springer.

\bibitem[Boyd and Pavlovski(2000)]{boyd2000attacking}
Colin Boyd and Chris Pavlovski.
\newblock Attacking and repairing batch verification schemes.
\newblock In \emph{Advances in Cryptology -- ASIACRYPT 2000}, volume 1976 of
  \emph{LNCS}, pages 58--71, Berlin, Heidelberg, 2000. Springer.

\bibitem[Hakuta et~al.(2013)Hakuta, Katoh, Sato, and Takagi]{hakuta2012batch}
Keisuke Hakuta, Yosuke Katoh, Hisayoshi Sato, and Tsuyoshi Takagi.
\newblock Batch verification suitable for efficiently verifying a limited
  number of signatures.
\newblock In \emph{Information Security and Cryptology -- ICISC 2012}, volume
  7839 of \emph{LNCS}, pages 425--440, Berlin, Heidelberg, 2013. Springer.

\bibitem[eip(2017)]{eip197}
{EIP-197}: Precompiled contracts for optimal ate pairing check on the elliptic
  curve alt\_bn128.
\newblock Ethereum Improvement Proposals,
  \url{https://eips.ethereum.org/EIPS/eip-197}, 2017.

\bibitem[{ConsenSys}(2023)]{goquorum}
{ConsenSys}.
\newblock {GoQuorum} documentation.
\newblock \url{https://docs.goquorum.consensys.io}, 2023.

\bibitem[Merkle(1988)]{merkle1987digital}
Ralph~C. Merkle.
\newblock A digital signature based on a conventional encryption function.
\newblock In \emph{Advances in Cryptology -- CRYPTO '87}, volume 293 of
  \emph{LNCS}, pages 369--378, Berlin, Heidelberg, 1988. Springer.

\bibitem[Miller(2025)]{noble}
Paul Miller.
\newblock noble-curves: Audited and minimal {JS} implementation of elliptic
  curve cryptography.
\newblock \url{https://github.com/paulmillr/noble-curves}, 2025.

\bibitem[Fiat and Shamir(1987)]{fiat1986prove}
Amos Fiat and Adi Shamir.
\newblock How to prove yourself: Practical solutions to identification and
  signature problems.
\newblock In \emph{Advances in Cryptology -- CRYPTO '86}, volume 263 of
  \emph{LNCS}, pages 186--194, Berlin, Heidelberg, 1987. Springer.

\bibitem[{EthereumJS contributors}(2025)]{ethereumjs}
{EthereumJS contributors}.
\newblock {EthereumJS} monorepo: {VM}, {EVM} and related libraries.
\newblock \url{https://github.com/ethereumjs/ethereumjs-monorepo}, 2025.

\bibitem[{Bitcoin Core contributors}(2025)]{libsecp256k1}
{Bitcoin Core contributors}.
\newblock libsecp256k1: Optimized {C} library for {ECDSA} signatures and
  secret/public key operations on curve secp256k1.
\newblock \url{https://github.com/bitcoin-core/secp256k1}, 2025.

\bibitem[{arkworks contributors}(2022)]{arkworks}
{arkworks contributors}.
\newblock arkworks: An ecosystem for {zkSNARK} programming.
\newblock \url{https://arkworks.rs}, 2022.

\bibitem[eip(2024)]{eip4844}
{EIP-4844}: Shard blob transactions.
\newblock Ethereum Improvement Proposals,
  \url{https://eips.ethereum.org/EIPS/eip-4844}, 2024.

\bibitem[eip(2025{\natexlab{h}})]{eip7918}
{EIP-7918}: Blob base fee bounded by execution cost.
\newblock Ethereum Improvement Proposals,
  \url{https://eips.ethereum.org/EIPS/eip-7918}, 2025{\natexlab{h}}.

\bibitem[eip(2025{\natexlab{i}})]{eip7594}
{EIP-7594}: {PeerDAS} -- peer data availability sampling.
\newblock Ethereum Improvement Proposals,
  \url{https://eips.ethereum.org/EIPS/eip-7594}, 2025{\natexlab{i}}.

\bibitem[Groth(2016)]{groth2016size}
Jens Groth.
\newblock On the size of pairing-based non-interactive arguments.
\newblock In \emph{Advances in Cryptology -- EUROCRYPT 2016}, volume 9666 of
  \emph{LNCS}, pages 305--326, Berlin, Heidelberg, 2016. Springer.
\newblock \doi{10.1007/978-3-662-49896-5_11}.

\bibitem[eip(2019)]{eip1108}
{EIP-1108}: Reduce alt\_bn128 precompile gas costs.
\newblock Ethereum Improvement Proposals,
  \url{https://eips.ethereum.org/EIPS/eip-1108}, 2019.

\bibitem[Al-Riyami and Paterson(2003)]{alriyami2003certificateless}
Sattam~S. Al-Riyami and Kenneth~G. Paterson.
\newblock Certificateless public key cryptography.
\newblock In \emph{Advances in Cryptology -- ASIACRYPT 2003}, volume 2894 of
  \emph{LNCS}, pages 452--473, Berlin, Heidelberg, 2003. Springer.

\bibitem[Galbraith et~al.(2008)Galbraith, Paterson, and
  Smart]{galbraith2008pairings}
Steven~D. Galbraith, Kenneth~G. Paterson, and Nigel~P. Smart.
\newblock Pairings for cryptographers.
\newblock \emph{Discrete Applied Mathematics}, 156\penalty0 (16):\penalty0
  3113--3121, 2008.

\bibitem[Pointcheval and Stern(2000)]{pointcheval2000security}
David Pointcheval and Jacques Stern.
\newblock Security arguments for digital signatures and blind signatures.
\newblock \emph{Journal of Cryptology}, 13\penalty0 (3):\penalty0 361--396,
  2000.

\end{thebibliography}

\end{document}